\documentstyle[seceq,letter,epsf]{ptptex}

\title{
Complex Energy Method for Scattering Processes
}

\author{
H. Kamada$^{1)}$
\footnote{email: kamada@mns.kyutech.ac.jp 
},
Y. Koike$^{2,3)}$ \footnote{email: koike@i.hosei.ac.jp},
W. Gl\"ockle $^{4)}$\footnote{email: walter.gloeckle@tp2.ruhr-uni-bochum.de}
}

\inst{
${}^{1)}$Department  of Physics, Faculty of Engineering, Kyushu Institute of Technology,
Kitakyushu 804-8550, Japan \\
${}^{2)}$
Science Research Center, Hosei University, 2-17-1, Fujimi, Chiyoda-ku, \\
Tokyo 102, Japan
\\
${}^{3)}$
Center for Nuclear Study, University of Tokyo, 2-1 Hirosawa, Wako, \\
Saitama 351-0198, Japan 
\\
${}^{4)}$
Institut f\"ur  Theoretische Physik II, Ruhr-Universit\"at Bochum, \\
D-44780 Bochum, Germany
}
\recdate{
\today
}

\abst{

A method for solving few-body scattering equations is proposed and examined.
The solution of the scattering equations at complex energies
is analytically continued to get scattering T-matrix with real positive energy.
Numerical examples document that  the method works well for  two-nucleon scattering and 
three-nucleon scattering, if the set of complex energies is properly chosen.

}

\begin{document}

\maketitle

\section{Introduction}

Scattering by more than two particles upon each other
requires boundary conditions 
in configuration space of
increasing complexity with  growing particle numbers. 
Those boundary conditions are reflected in
momentum space by $\epsilon \to$ 0 limits of different 
resolvent operators $G(E+i\epsilon )$,
where E is real and above some thresholds. While such
a singularity can easily be
handled on a two-body fragmentation cut,  it causes 
already severe difficulties on a
three-body fragmentation cut like in the three-body
problem. There it leads to so called
moving singularities\cite{S-Z,Hueber-Kamada}.
 They can be treated directly for 
instance using spline interpolations
or just straightforward subtraction methods\cite{Gloeckle3}. They can
also be circumvented by contour
deformations of the path of  momentum integration\cite{Lovalace,Karhill,Ehben,Koike-d-a}.
In any case this requires great care.
For more than three particles the
four-body break
up singularities pose even harder challenges.

In the past several additional  methods have been
proposed especially for two-body
fragmentations. For instance in \cite{Schlessinger} 
the T-
matrix is evaluated at negative energies , where bound
state methods can be employed, and
then analytically continued to the energy  in the
scattering region. In \cite{McDonald,Reinhardt,Doolen} 
the T-matrix is evaluated at several complex energies
above the positive real energy axis
and then the value on the real energy axis is again 
determined by analytical continuation. All the applications
we are aware of are in the
field of atomic physics.

The Lorentz transformation method used intensively by 
the Trento group \cite{Trento}  solves many
body equations based on a complex energy, thereby 
avoiding the complicated boundary
conditions in configuration space or when used in
momentum space \cite{Martinelli} avoiding
singularities. But in this case no continuation to
the real energy axis is performed and
the  imaginary part of the energy is kept fixed.
Certain inversion procedures are
subsequently employed.

  Here we would like to reconsider the approach
evaluating the two- and three-body T-matrix for
complex energies and performing an analytical 
continuation to the real axis. In view of
future applications to the field of nuclear physics
we shall employ nucleon - nucleon
forces .  
Section \ref{sec2} displays our studies for two nucleons.
We add  a brief example for  three-body scattering
in section \ref{sec3} end with a brief outlook in section 4.

 \section{Two- Nucleon Scattering}\label{sec2}

We start with a simple model for the NN forces
allowing for an analytical solution, the
Yamaguchi separable force \cite{Yamaguchi}
\begin{eqnarray}
v(p,p')&=&\lambda { 1 \over {p^2 + \beta ^2}}{ 1 \over {{p'} ^2 + \beta
^2  }}
\end{eqnarray}         
It leads to the two-body T- matrix
\begin{eqnarray}
t (p,p'; E) &=& \tau (E){ 1 \over {p^2 + \beta ^2}}{ 1 \over {{p'} ^2 + \beta
^2  }} 
\end{eqnarray}    
with
\begin{eqnarray}
\tau (E) &=& (  \lambda ^{-1} + { \pi (\beta ^2 - k^2 ) m \over 4 \beta (\beta^2 +k^2)^2 }
+ i { \pi k m \over 2 (\beta ^2 +k^2 )^2 }  )^{-1}
\label{eq.tau}
\end{eqnarray}  
Now we evaluate $\tau(E)$ replacing energy
$E\equiv k^2/m $ by $E+i \epsilon$,
where $\epsilon$ is now finite and we choose
a set of discrete $\epsilon$'s. That function $\tilde \tau (\epsilon) \equiv 
\tau (E+i\epsilon) $ can be analytically
continued to $\epsilon =$0. 
Pad\'e approximant is well known as a method for the analytical continuation.  It
is useful when coefficients of the Taylor series are given.  On the contrary, the
functional values at several points are given in the present case. 
We employ the point method \cite{Schlessinger} and  consider the continued fraction 
\begin{eqnarray}
\tilde \tau (\epsilon ) &=& { \tau (E + \epsilon_1 i) \over { 1 +
{ a_1 (\epsilon -  \epsilon_1 ) \over { 1 + \cdots  }  } } }
\cr
&=& { \tau (E + \epsilon_1 i) \over { 1 +} }
{ a_1 (\epsilon -  \epsilon_1 ) \over { 1 +} }
{ a_2 (\epsilon -  \epsilon_2 ) \over { 1 +} } \cdots  
\label{eq.point}
\end{eqnarray}
The coefficients  $a_i$ are easily deduced from 
the discrete set of $\tilde \tau  ( \epsilon )$-values:
\begin{eqnarray}
a_l &=& { 1 \over { \epsilon_l - \epsilon_{l+1}}} 
\{ 1 + { a_{l-1} (\epsilon_{l+1} -  \epsilon_{l-1} ) \over { 1 +} }
{ a_{l-2} (\epsilon_{l+1} -  \epsilon_{l-2} ) \over { 1 +} }
\cdots \cr 
&+& { a_1 (\epsilon_{l+1} -  \epsilon_{1} ) \over { 1 - 
[ \tau (E +i \epsilon_1) / \tau (E+i\epsilon_{l+1} ) ]  } }
\}
\end{eqnarray}
and 
\begin{eqnarray}
a_1 &=&  { [ \tau (E +i \epsilon_1) / \tau (E+i\epsilon_{2} ) ] -1 \over 
{ \epsilon_2 - \epsilon_{1} }  }.
\end{eqnarray}
An example is displayed in Table I, which shows 
the smooth behavior of $\tilde \tau  ( \epsilon)$
sampled for discrete
$\epsilon$- values according to Eq(\ref{eq.tau}).
The analytically continued value
for $\epsilon$ =0 as given by (\ref{eq.point}) agrees perfectly well with the exact one.

\begin{table}[tp]
\begin{center}
\begin{tabular}[t]{l|l|}
 $\epsilon$ [MeV] & $\tau (E + i \epsilon )$ [fm$^2$] \cr
\hline
0.1  &-0.11701981630329802~~-0.60857451562152798 $i$ \cr
0.2  &-0.11924707235802381~~-0.61103993137637069 $i$ \cr
0.3  &-0.12151390252710244~~-0.61348507952563247 $i$ \cr
0.4  &-0.1238200220019596~~~-0.6159089815519192   $i$ \cr
0.5  &-0.12616511343686601~~-0.61831066360816722 $i$ \cr
0.6  &-0.12854882700065837~~-0.62068915788211088 $i$ \cr
0.7  &-0.13097078049779975~~-0.62304350396337027 $i$ \cr
0.8  &-0.13343055955913999~~-0.6253727502090104  $i$ \cr
0.9  &-0.13592771790246552~~-0.62767595510338259 $i$  \cr
1.0  &-0.13846177766262588~~-0.62995218860804847 $i$ \cr
\hline
0     &-0.1148323866513037~~~-0.60608981409107898 $i$ \cr
exact &-0.11483238665130231~~-0.60608981409107832 
$i$ \cr
   \end{tabular}
  \caption{The propagator $\tau$
 from Eq(\ref{eq.tau}) for different $\epsilon$
- values. The value for $\epsilon$=0 is evaluated by
the point method \cite{Schlessinger}. The exact value is obtained from Eq.(\ref{eq.tau}). 
The parameters are $\lambda$=-0.5592[fm$^2$],$\beta$=1.13[fm$^{-1}$] and
$m^{-1}$=41.47[MeVfm$^2]$. The energy is 10.0 MeV.}
   \label{table1}
 \end{center}
\end{table}

\begin{table}[tp]
\begin{center}
\begin{tabular}[t]{l|l|l|l|}
$\epsilon$ [MeV] & $T _{0 0} $  [fm$^2$] & $T _{0 2}$   [fm$^2$] & $T _{2 2}$   [fm$^2$] \cr
\hline
0.4  & 0.074878-0.38063 $i$ & -9.7683 $\times 10^{-4}$+8.8418$\times 10^{-3}$ $i$ & 4.5582$\times 10^{-3}$-2.9901$\times 10^{-4}$ $i$ \cr
0.5  & 0.072896-0.38404 $i$ & -8.0207$ \times 10^{-4}$+9.1311$\times 10^{-3}$ $i$ & 4.5523$\times 10^{-3}$-3.1989$\times 10^{-4} $$i$ \cr
0.6  & 0.070821-0.38742 $i$ & -6.2122 $\times 10^{-4}$+9.4171$\times 10^{-3}$ $i$ & 4.5460$\times 10^{-3}$-3.4052$\times 10^{-4}$$i$\cr
0.7  & 0.068651-0.39077 $i$ & -4.3437 $\times 10^{-4}$+9.6996$\times 10^{-3}$ $i$ & 4.5392$\times 10^{-3}$-3.6090$\times 10^{-4}$ $i$\cr
0.8  & 0.066387-0.39409 $i$ & -2.4160 $\times 10^{-4}$+9.9782$\times 10^{-3}$ $i$ & 4.5319$\times 10^{-3}$-3.8100$\times 10^{-4}$ $i$\cr
\hline
0  & 0.081860-0.36687 $i$ & -1.6137  $\times 10^{-3}$+7.6562 $\times 10^{-3}$ $i$ & 4.5768 $\times 10^{-3}$-2.1337 $\times 10^{-4}$ $i$\cr
exact  & 0.081860-0.36687 $i$ & -1.6138  $\times 10^{-3}$+7.6564 $\times 10^{-3}$ $i$ & 4.5767 $\times 10^{-3}$-2.1338 $\times 10^{-4}$ $i$\cr
   \end{tabular}
  \caption{ The T-matrix for the  $^3$S$_1$-$^3$D$_1$ state and different
$\epsilon$-values. 
The value for $\epsilon$=0 is calculated by
the point method and compared to  the exact number resulting directly
from
Eq.(\ref{LS.eq}).   }
   \label{table2}
 \end{center}
\end{table}

Now that approach also works perfectly for a 
modern high precision NN force like the  CD-Bonn potential
\cite{CDBONN}. As an example we take the coupled channel
case $^3S_1-^3D_1$ and solve now the two-
body Lippmann Schwinger equation 
\begin{eqnarray}
T _{l' l } (p' ,p ) = V_{l' l} (p' ,p ) +
\sum_{l''}\int _0 ^ \infty dp'' {p''}^2  V_{l' l''} (p' ,p'' )
{ 1 \over { E+i \epsilon  -{{p''}^2\over m} }  }   T_{l'' l} (p'' ,p
)~~~~ 
\label{LS.eq}
\end{eqnarray}
numerically for a set of $\epsilon$- values. We use
the same method.
The continued fraction  is now in the
matrix
\begin{eqnarray}
\tilde T ( \epsilon) \equiv T (p,p; E+ i\epsilon) =  
{ T (p,p;E + \epsilon_1 i) \over { 1 +} }
{ A_1 (\epsilon -  \epsilon_1 ) \over { 1 +} }
{ A_2 (\epsilon -  \epsilon_2 ) \over { 1 +} } \cdots 
\end{eqnarray}
where $A_i$ are components of 2$\times$2 matrices.
We arrive at the results displayed in Table \ref{table2}.
There we show the on-
shell T- matrix elements $T_{00}(p,p),T_{02}(p,p), T_{22}(p,p)$ 
for $p=\sqrt{m E}$, $E=$10.0 MeV and different
$\epsilon$-values. Again we see a smooth 
dependence on $\epsilon$.
The resulting matrix from the point method elements ($\epsilon$=0)
are compared
to the values evaluated by a standard
subtraction method in Table  \ref{table2}. The agreement 
is again perfect.

That expansion works equally well for off- shell 
T- matrices $\lim_{\epsilon \to 0 } T( p', p; E+i \epsilon  )$
where $p \ne p'$  and $E$  not
connected to $p$ or $p'$.

\section{An example for three-nucleon scattering}
\label{sec3}

We now proceed to  three-nucleon scattering.  Several methods have been 
developed to treat singularities for the Faddeev equations.  
The contour deformation method has been used as one of the methods for avoiding 
complicated singularities.  
Here we adopt a separable expansion method \cite{Koike-sep1,Koike-sep2}
for calculating the two-body t-matrix. The resulting
three-body equations become coupled integral equations in one variable and the
contour deformation can be used as a tool for  avoiding singularities.  We compare
the three-nucleon scattering amplitudes obtained by the present method with those
from the traditional contour deformation method.

The three-body equations we solve are of the following form
\begin{eqnarray}
  X^{(J^\pi)}_{cc'}&&(q,q';E) =  Z^{(J^\pi)}_{cc'}(q,q';E)    \cr
    &&+ \sum_{c''} \int_0^\infty
    q''^2 dq'' Z^{(J^\pi)}_{cc''}(q,q'';E) \tau(E-q''^2/2\mu)
     X^{(J^\pi)}_{c''c'}(q'',q';E)  \label{AL}
\end{eqnarray}
Eqs. (\ref{AL}) are a generalization of the Amado-Mitra-Faddeev-Lovelace(AMFL) equations
, which are three-body equations based on a  simple rank-1 separable potential.  The
AMFL equations are described in  standard textbooks for the quantum mechanical three-body
problem and the generalization thereof in ref\cite{Parke}.

The three-body amplitude describing  elastic scattering is the on-shell n-d amplitude
$X^{(J^\pi)}_{c_ic_j}(q_0,q_0;E)$ with $E = q_0^2 / 2\mu$ and  n-d channels $c_i,c_j$ 
allowed for the total spin and parity $J^\pi$.  The number of  
n-d channels are two for $J^\pi = (1/2)^+$, $(1/2)^-$ or three for others.  The contour deformation utilizes 
the analytical continuation in the  momenta $q, q', q''$, while the present method is based 
on 
the analytical continuation in  the energy $E$.  In both cases,the  various singularities in
Eq.(\ref{AL}) are avoided.   

We solve equ(3.1)  with the present method at 10 MeV laboratory energy.
At this energy, it is known that the breakup cross
section is already about 10-20 \% of the total cross section.  Thus the moving 
singularity is quite  important.

\begin{table}[tp]
\begin{center}
\begin{tabular}[t]{l|l|l|l|}
 $i \epsilon$ [MeV] & $X _{d d}^{(J^\pi)} $  [fm$^5$] & $X_{d q}^{(J^\pi)}$   [fm$^5$] & $X_{q q}^{(J^\pi)}$   [fm$^5$] \cr
\hline
 0+1.0 $i$  &  0.5634 -1.7745 $i$ &-1.3755  $\times 10^{-2}$+1.4227$\times 10^{-2}$ $i$ & 0.2618+4.5351$\times 10^{-2}$ $i$ \cr
1.0+1.5 $i$  & 0.6669 -1.8203 $i$ & -1.9726 $\times 10^{-2}$+8.4048$\times 10^{-3}$ $i$ & 0.2451+8.7036$\times 10^{-2}$ $i$ \cr
0.0+2.0 $i$  & 0.7608 -1.8267 $i$ & -2.5101 $\times 10^{-2}$+2.3236$\times 10^{-3}$ $i$ & 0.2198+0.1208 $i$\cr
0.5+1.5 $i$  & 0.6513 -1.9441 $i$ & -2.5979 $\times 10^{-2}$+1.4182$\times 10^{-2}$ $i$ & 0.2878+0.1136 $i$\cr
-0.5+1.5 $i$ & 0.6937 -1.7363 $i$ & -1.4067 $\times 10^{-2}$+2.8661$\times 10^{-3}$ $i$ & 0.2109+6.8861$\times 10^{-2}$ $i$\cr
 0.5+1.0 $i$ & 0.4892 -1.9070 $i$ & -1.9834 $\times 10^{-2}$+2.0626$\times 10^{-2}$ $i$ & 0.3135 +5.9376$\times 10^{-2}$ $i$\cr
-0.5+1.0 $i$ & 0.6265 -1.6940 $i$ & -8.3854 $\times 10^{-2}$+8.0987$\times 10^{-3}$ $i$ & 0.2225+ 3.6096$\times 10^{-2}$ $i$\cr
 0.5+2.0 $i$ & 0.7819 -1.9252 $i$ & -3.1597 $\times 10^{-2}$+7.5166$\times 10^{-3}$ $i$ &0.2514+0.1543 $i$\cr
\hline
 0   & 0.4248 -1.5440 $i$ & 2.5154 $\times 10^{-5}$+ 2.4663$\times 10^{-2}$ $i$ & 0.2586 -4.5922 $\times 10^{-2}$ $i$\cr
C. D.& 0.4244 -1.5441 $i$ & 4.9443 $\times 10^{-5}$+ 2.4784$\times
10^{-2}$ $i$ & 0.2585 -4.5875$\times 10^{-2}$ $i$\cr
    \end{tabular}
   \caption{T-matrix for three-body scattering  for different
complex $\epsilon$-values and  $J^\pi$ = (1/2)$^+$.  
The value on the real axis ($i \epsilon$=0)  is calculated by
the point method and compared to  the  contour deformation number (C. D.) resulting directly
from  Eq.(\ref{AL}). The channels $d$ and $q$ correspond to 
spin doublet and quartet, respectively.}
    \label{table3}
  \end{center}
\end{table}

\begin{table}[tp]
\begin{center}
\begin{tabular}[t]{l|l|l|l|}
 $i \epsilon$ [MeV] & $X _{d d}^{(J^\pi)} $  [fm$^5$] & $X_{d q}^{(J^\pi)}$   [fm$^5$] & $X_{q q}^{(J^\pi)}$   [fm$^5$] \cr
\hline
 0+1.0 $i$ & 0.2117 -0.3215 $i$   & -0.2277 -0.2446 $i$ &  -0.6525 -0.6931  $i$ \cr
1.0+1.5 $i$  & 0.3032 -0.2590 $i$ & -0.1941 -0.2410 $i$ &  -0.5818 -0.6991 $i$ \cr
0.0+2.0 $i$  & 0.3575 -0.1809 $i$ & -0.1650 -0.2328 $i$ &  -0.5161 -0.6986 $i$\cr
0.5+1.5 $i$  & 0.3999 -0.3379 $i$ & -0.1851 -0.2748 $i$ &  -0.5806 -0.7702 $i$\cr
-0.5+1.5 $i$ & 0.2454 -0.1941 $i$ & -0.1984 -0.2116 $i$ &  -0.5775 -0.6340 $i$\cr
 0.5+1.0 $i$ & 0.2813 -0.4530 $i$ & -0.2250 -0.2838 $i$ &  -0.6601 -0.7722 $i$\cr
-0.5+1.0 $i$ & 0.1778 -0.2291 $i$ & -0.2271 -0.2121 $i$ &  -0.6406 -0.6249 $i$\cr
 0.5+2.0 $i$ & 0.4549 -0.2156 $i$ & -0.1526 -0.2606 $i$ &  -0.5111 -0.7635 $i$\cr
\hline
0   & -4.5247$\times 10^{-2}$ -0.3098 $i$ & -0.3013 -0.2316 $i$ & -0.7976 -0.6440 $i$\cr
C. D.&-4.5393$\times 10^{-2}$ -0.3098  $i$ & -0.3016 -0.2312 $i$ & -0.7975 -0.6437 $i$\cr
    \end{tabular}
   \caption{Same as table \ref{table3} except for  $J^\pi$ = (1/2)$^-$.
  }
    \label{table4}
  \end{center}
\end{table}

In Table \ref{table3}, we show a  result for the three-nucleon scattering amplitude 
for $J^\pi = (1/2)^+$ based on  the AV14 potential\cite{AV14}.  
Two channels, denoted by d and q and reflecting spin states,  are coupled for
 the n-d elastic scattering amplitude.
The energy is chosen as  $E = E_0 + i \epsilon$, where 
$E_0 = q_0^2/2\mu$ and $q_0$ is  the initial momentum. In contrast to the simpler
two-body singularity treated in section 2 , where real epsilons were chosen, we use now
also  complex epsilons. This allows to have more complex energy points in the
neighborhood of $E_0$. Also we kept the imaginary parts of the energy always larger than 1
MeV ( this is in relation to the chosen $E_0$=10 MeV) which guaranteees a relatively smooth
integrand and avoids problems in the convergence of the integration. As a consequence of
that large imaginary parts of the discrete set of energies the amplitudes shown in Table 3
differ considerably from the amplitudes, C.D., on the real axis as obtained with the contour
deformation method. Nevertheless the point method applied to those amplitudes leads to
good converged results which are very close to the amplitudes C.D.

Table \ref{table4} shows the amplitudes for $J^\pi = (1/2)^-$.  We see again the same good
convergence to the results achieved by the contour deformation method.
  
\section{Outlook}
We demonstrated the easiness by which the singularities
in the Lippmann Schwinger equation 
for NN scattering can be avoided. We used a set
of complex energies above the positive
real axis and based on that performed an analytical 
continuation to the real axis. 
This analytical continuation was performed by the point
method ( a continued fraction expansion). 

In case of the three-body Faddeev equations one
encounters in the integral kernel two types of
singularities related to the nucleon-deuteron and
three-nucleon cuts.
The first one is
generated by the off- shell NN T- matrix and the 
second one by the free three-nucleon
propagator.
As shown in section \ref{sec3} the complex energy method 
can again be 
applied.
The resulting accuracy is quite satisfactory and the resulting amplitudes
 agree very well
with the ones from the contour deformation method.

This method also  allows
to search for resonances below the real axis.
In a forthcoming paper 
we will display that this  method is applicable 
to even more than three particles like the
complex four- body break- up process , where no
calculations have been performed so far.

\section*{Acknowledgements}
One of authors (H.K.) is grateful to I. Fachruddin, A. Hemmdan,
E. Epelbaum,
R. Skibi\'nski and H. H. Oo not only for numerous discussions but also for 
the kind  atmosphere when he recently visited 
the Ruhr-Universit\"at Bochum.


\begin{thebibliography}{99}

\bibitem{S-Z} E. W. Schmid and H. Ziegelmann, 
"The Quantum Mechanical Three-Body Problem"(Pergamon Press, Oxford 1974)

\bibitem{Hueber-Kamada} D. H\"uber, H. Kamada, H. Wita\l a,
W. Gl\"ockle, Few-Body Systems {\bf 16}, 165 (1994).


\bibitem{Gloeckle3} W. Gl\"ockle, H. Wita\l a, D. H\"uber, H. Kamada,
and J. Golak,  Phys. Rep. {\bf  274}, 107 (1996).



\bibitem{Lovalace} C. Lovelace, Phys. Rev C {\bf 135 }, 1125 (1964).


\bibitem{Karhill} R. T. Cahill, I. H. Sloan, Nucl. Phys. {\bf A165},
161 (1971).
                                                    
\bibitem{Ehben}   W. Ebenh\"oh, Nucl. Phys. {\bf A191}, 97 (1972).

\bibitem{Koike-d-a} Y. Koike, Nucl. Phys. {\bf A301}, 411 (1978).

\bibitem{Schlessinger}  L. Schlessinger,
Phys. Rev. {\bf 167}, 1411 (1968).


\bibitem{McDonald}  F. A. McDonald,
J.Nuttall, Phys. Rev. A {\bf 4}, 1821 (1971).

\bibitem{Reinhardt} W. P. Reinhardt,
D. W. Oxtoby, T. N. Resicgno, Phys. Rev.
Lett. {\bf 28}, 401 (1972).

\bibitem{Doolen}  G. Doolen, G. McCartor, F. A. McDonald,
J. Nuttall Phys. Rev A {\bf 4}, 108 (1971).

\bibitem{Trento} V. D. Efros, W. Leidemann, G. Orlandini, Phys. Lett. {B} {\bf 338}, 130 (1994); V. D. Efros,
Yad. Fiz. {\bf 41}, 1498 (1985) [Sov. J. Nucl. Phys. {\bf 41}, 949 (1985)].   

\bibitem{Martinelli} S. Martinelli, H. Kamada, G. Orlandini, and
W. Gl\"ockle, Phys. Rev. C {\bf 52}, 2906 (1995).


\bibitem{Yamaguchi} Y. Yamaguchi, Phys. Rev {\bf 95 }, 1628 (1954).     


\bibitem{CDBONN} 
R. Machleidt, F. Sammarruca, and Y. Song, Phy. Rev. C{\bf 53}, R1483 (1996).

\bibitem{Faddeev} L. D. Faddeev, Sov. Phys. JETP {\bf 12  }, 1014
(1961).

\bibitem{AGS} E. O. Alt, P. Grassberger, W. Sandhas, Nucl. Phys. {\bf
B2 }, 167 (1967).



\bibitem{Gloeckle} W. Gl\"ockle, {\it The Quantum-Mechanical Few-Body Problem} (
Springer
Verlag, Berlin, Heiderberg, 1983).          


\bibitem{Koike-sep1} Y. Koike, Phys. Rev. C {\bf 42}, R2286 (1990).

\bibitem{Koike-sep2} Y. Koike, W. C. Parke, L. C. Maxmon, and D.R. Lehman,
Few-Body Systems {\bf 23}, 53 (1997).

\bibitem{Parke} W. C. Parke, Y. Koike, D. R. Lehman, and L. C. Maxmon,
Few-Body Systems {\bf 11}, 89 (1991).

\bibitem{AV14} R. B. Wiringa, R. A. Smith, and T. L. Ainsworth,
Phys. Rev. C {\bf 29}, 1207 (1984).

\end{thebibliography}
\end{document}